\documentclass[%
reprint,
superscriptaddress,
 amsmath,amssymb,
 aps,
prb,
floatfix,
]{revtex4-1}

\usepackage{graphicx}
\usepackage{dcolumn}
\usepackage{bm}

\usepackage{gensymb}	

\usepackage[bookmarks=true,colorlinks=true,linkcolor={blue},citecolor=blue,urlcolor=blue]{hyperref} 
\usepackage[mathlines]{lineno}

\usepackage{longtable}	
\usepackage{tabularx}	
\usepackage{array}		
\usepackage{multirow}	
\usepackage{array}		
\begin{document}


\title{Quantum mechanical \textit{ab initio} calculations of the structural, electronic and optical properties of bulk gold nitrides}


\author{Mohammed S. H. Suleiman}
\email[Corresponding author: ]{suleiman@aims.ac.za}
\affiliation{School of Physics, University of the Witwatersrand, Johannesburg, South Africa.}
\affiliation{Department of Physics, Sudan University of Science and Technology, Khartoum, Sudan.}

\author{Daniel P. Joubert}
\homepage[Homepage: ]{http://www.wits.ac.za/staff/daniel.joubert2.htm}
\affiliation{School of Physics, University of the Witwatersrand, Johannesburg, South Africa.}

\date{\today}

\begin{abstract}
In the present work, the atomic and the electronic structures of Au$_{3}$N, AuN and AuN$_{2}$ are investigated using first-principles density-functional theory (DFT). We studied cohesive energy \textit{vs.} volume data for a wide range of possible structures of these nitrides. Obtained data was fitted to a Birch-Murnaghan
third-order equation of state (EOS) so as to identify the most likely candidates for the true crystal
structure in this subset of the infinite parameter space, and to determine their equilibrium structural
parameters. The analysis of the electronic properties was achieved by the calculations of the band 
structure and the total and partial density of states (DOS). Some possible pressure-induced
structural phase transitions have been pointed out. Further, we carried out GW$_{0}$ calculations within the random-phase approximation (RPA) to the dielectric tensor to investigate the optical spectra of the experimentally suggested modification: Au$_{3}$N(D0$_{9}$). Obtained results are compared with experiment and with some available previous calculations.
\end{abstract}


\maketitle

\tableofcontents	
%
\section{Introduction} \label{Introduction}
In 2002, \ifmmode \check{S} \else \v{S}\fi{}iller and co-workers \cite{1st_gold_nitride_2002}$^{,}$
\footnote{In fact, the most noticeable works are those of \v{S}iller and co-workers \cite{1st_gold_nitride_2002, PhysRevB.70.045414, gold_nitride_2005_expt, gold_nitride_2006_expt, gold_nitrides_2008_expt, gold_ntride_2009_expt, gold_nitride_review_2009} at the University of Newcastle. See \url{http://research.ncl.ac.uk/nanoscale/research/goldnitride.html}, \url{http://news.bbc.co.uk/2/hi/uk_news/england/tyne/3205959.stm} and \url{http://physicsworld.com/cws/article/news/2003/oct/27/gold-plating-on-the-cheap}.}
 reported direct observation of the formation of an Au$_{x}$N compound for the first time. Since then, single crystal and polycrystalline gold nitrides have been prepared with different methods \cite{gold_nitrides_2008_expt,gold_nitride_review_2009}, and many theoretical \cite{PhysRevB.70.045414,AgN2_AuN2_PtN2_2005_comp,CuN_AgN_AuN_2007_comp,AuN2_and_PtN2_2010_comp} and experimental \cite{PhysRevB.70.045414,gold_nitride_2005_expt,gold_nitrides_2008_expt,AuN_2008_expt,gold_ntride_2009_expt,gold_nitride_2006_expt} investigations on the structural and physical properties of gold nitride have been published. It turned out that gold nitride possesses interesting properties which may lead to potential practical applications \cite{gold_nitride_2006_expt}.

So far, the most significant finding may be that of \ifmmode \check{S} \else \v{S}\fi{}iller et. al \cite{gold_nitride_2005_expt} who, in 2005, reported the production of metallic large area gold nitride films which are $\sim 50 \%$ harder than pure gold films produced under similar conditions,
making the gold nitride ideal for use in large-scale applications in coatings and in electronics. Moreover, the possibility of patterning gold nitride film surfaces by electron/photon beam lithography was confirmed \cite{gold_nitride_2006_expt}.

From their experimental observations and \textit{ab initio} calculations, Krishnamurthy \textit{et al.} \cite{PhysRevB.70.045414} suggested the possibility of formation of more than one gold nitride phase. Although theoretical calculations have predicted several possible structures for AuN, AuN$_{2}$ and Au$_{3}$N, none of these agrees with experiment \cite{gold_nitrides_2008_expt}.\\

To identify the most likely candidates for the true stoichiometry and the true crystal structure, we investigate the structure preference and thermodynamic stability of gold nitride in three different chemical formulas: Au$_3$N, AuN, and AuN$_2$ in $20$ structural modifications. The electronic properties of the most stable candidate in each stoichiometric series, and the optical properties of a previously suggested modification, Au$_{3}$N(D0$_{9}$), are carefully studied.
%
\section{Calculation Details} \label{Calculation Details}

\subsection{Crystal Structures and Chemical Formulae}	\label{Crystal Structures and Chemical Formulae}
Information about the considered crystal structures are given in Table \ref{AuN's allstructures} below. In this table, structures are first grouped according to the nitrogen content, starting with the stoichiometry with the lowest nitrogen content Au$_3$N, followed by the 1:1 series and ending with the nitrogen-richest AuN$_2$ group. Within each series, structures are ordered according to their structural symmetry, starting from the highest symmetry (i.e. the highest space group number) to the least symmetry.
\begin{table}[h]
\caption{The studied structural phases of Au, Au${_3}$N, AuN and AuN${_2}$. Presented are the Strukturbericht symbol, the underlying Bravais lattice (BL), prototype compounds, the space group (SG), and the number of Au${_\text{m}}$N${_\text{n}}$ formulas per unit cell ($Z$).}
\resizebox{0.49\textwidth}{!}{
\begin{tabular}{l|llll}
\hline \hline	
Symbol			&BL				&Prototype(s)		&SG			&$Z$      \\
\hline
\multicolumn{5}{c}{\textbf{Au Structure}} \\
\hline	
A1              &fcc            	&Cu				&Fm$\bar{3}$m   & 1       \\

\hline	
\multicolumn{5}{c}{\textbf{Au$_3$N Structures}} \\
\hline	
D0$_3$            &fcc                		&AlFe$_3$					&Fm$\bar{3}$m 	&1 		\\

A15               &sc                       &Cr$_3$Si 					&Pm$\bar{3}$n	&2     		\\

D0$_9$            &sc                       &anti-ReO$_3$ ($\alpha$), Cu$_3$N	&Pm$\bar{3}$m	&1     \\

L1$_2$            &sc                       &Cu$_3$Au                   &Pm$\bar{3}$m  		&1     	\\

D0$_2$            &bcc                      &CoAs$_3$ (skutterudite)    &Im$\bar{3}$   		&4     	\\

$\epsilon$-Fe$_3$N&hexagonal		        &$\epsilon$-Fe$_3$N, Ni$_3$N 	&P6$_{3}$22         &2     \\

RhF$_3$          &trigonal (rhomboedric) 	&RhF$_3$                    &R$\bar{3}$c   		&2     	\\


\hline	
\multicolumn{5}{c}{\textbf{AuN Structures}} \\
\hline	
B1              &fcc            	&NaCl				&Fm$\bar{3}$m   & 1       \\

B2              &sc             	&CsCl				&Pm$\bar{3}$m   & 1       \\  

B3              &fcc            	&ZnS (zincblende)	&F$\bar{4}3$m   & 1       \\   

B8$_{1}$        &hexagonal      	&NiAs				&P$6_{3}$/mmc   & 2       \\

B$_{\text{k}}$  &hexagonal      	&BN					&P$6_{3}$/mmc   & 2       \\

B$_{\text{h}}$  &hexagonal      	&WC					&P$\bar{6}$m$2$ & 1       \\

B4              &hexagonal      	&ZnS (wurtzite)		&P$6_{3}$mc     & 2       \\

B17             &s tetragonal   	&PtS (cooperite)	&P$4_{2}$/mmc   & 2       \\

B24             &fc orthorhombic	&TlF				&Fmmm            & 1       \\

\hline	
\multicolumn{5}{c}{\textbf{AuN$_2$ Structures}} \\
\hline	
C1			&fcc			&CaF$_{2}$ (fluorite)	&Fm$\bar{3}$m   & 1       \\
C2			&sc				&FeS$_{2}$ (pyrite)		&Pa$\bar{3}$    & 4       \\  

C18			&s orthorhombic	&FeS$_{2}$ (marcasite)	&Pnnm			& 2       \\   

CoSb$_{2}$	&s monoclinc	&CoSb$_{2}$	 			&P2$_1$/c		& 4       \\

\hline\hline
\end{tabular}
}	
\label{AuN's allstructures}
\end{table}
%
\subsection{DFT Calculation Details}	\label{DFT Calculation Details}
VASP code\cite{Vasp_ref_PhysRevB.47.558_1993,Vasp_ref_PhysRevB.49.14251_1994,Vasp_cite_Kressw_1996,
Vasp_PWs_Kresse_1996,DFT_VASP_Hafner_2008,PAW_Kresse_n_Joubert} was used for electronic structure spin density functional theory (SDFT) \cite{SDFT_1972,SDFT_Pant_1972} calculations. Here, a projector augmented wave (PAW) \cite{PAW_Blochl, PAW_Kresse_n_Joubert} description of the ion-electron interaction $V_{ext}(\mathbf{r})$ is implemented, where the $2s^{2}2p^{3}$  electrons of N and  the $5d^{10}6s^{1}$ electrons of Au are treated explicitly. While the PAW potential treats the core electrons in a fully relativistic fashion\cite{DFT_VASP_Hafner_2008}, only scalar kinematic relativistic effects for these valence electrons are incorporated. Spin-orbit interactions of the valence electrons have not been considered.

With $i$, $\mathbf{k}$ and $\sigma$ being the band, $\mathbf{k}$-point and spin indices, respectively, the Kohn-Sham (KS) equations \cite{KS_1965}
\begin{eqnarray}	\label{KS equations}
\begin{split}
  \Bigg \{ - \frac{\hbar^{2}} {2m_{e}}  \nabla^{2} + \int d\mathbf{r}^{\prime} \frac{n(\mathbf{r}^{\prime})}{|\mathbf{r}-\mathbf{r}^{\prime}|} + V_{ext}(\mathbf{r}) \\  + V_{XC}^{\sigma, \mathbf{k}}[n(\mathbf{r})] \Bigg \} \psi_{i}^{\sigma, \mathbf{k}}(\mathbf{r})  =  
   \epsilon_{i}^{\sigma, \mathbf{k}} \psi_{i}^{\sigma, \mathbf{k}}(\mathbf{r}),
\end{split}
\end{eqnarray}
are solved by expanding $\psi_{i}^{\sigma , \mathbf{k}}(\mathbf{r})$, the pseudo part of the KS one-particle spin orbitals, on a basis set of plane-waves (PWs). It is found that the total energy converged to less than $3 \; \text{m} eV/ \text{atom}$ using cut-off energy $E_{\text{cut}} \leq 600 \; eV$ and $\mathbf{\Gamma}$-centered Monkhorst-Pack \cite{MP_k_mesh_1976} $17 \times 17 \times 17$ meshes for the Brillouin zones (BZs) sampling. In the ionic relaxation stage, partial occupancies were set using the smearing method of Methfessel-Paxton (MP) \cite{MP_smearing_1989} and Fermi surface of the metallic phases has been carefully treated, while the tetrahedron method with Bl\"{o}chl corrections \cite{tetrahedron_method_theory_1971,tetrahedron_method_theory_1972,ISMEAR5_1994} was used in the static calculations. The Perdew-Burke-Ernzerhof (PBE) \cite{PBE_GGA_1996,PBE_GGA_Erratum_1997,XC_PBE_1999} GGA \cite{XC_GGA_1988,XC_GGA_applications_1992,XC_GGA_applications_1992_ERRATUM} exchange-correlation functional $V_{XC}^{\sigma, \mathbf{k}}[n(\mathbf{r})]$ was employed.
%
\subsection{Structural Relaxation and EOS} \label{Structural Relaxation}
To optimize the geometry, those atoms which possess internal free parameters were allowed to move till all Hellmann-Feynman force components \cite{Hellmann–Feynman_theorem} on each ion were  $< 1 \times 10^{-2} \; eV/\text{\AA}$; then static total energy calculation (as described in Subsection \ref{DFT Calculation Details}) followed. This was done at a set of isotropically varying volumes of the unit cells, and cohesive energy per atom \cite{Grimvall,Suleiman_PhD_arxiv_copper_nitrides_article}
\begin{eqnarray} \label{E_coh equation}
E_{\text{coh}}^{\text{Au}_{m}\text{N}_{n}}  =   \frac{  E_{\text{solid}}^{\text{Au}_{m}\text{N}_{n}} - Z \times \left( m E_{\text{atom}}^{\text{Au}} + n E_{\text{atom}}^{\text{N}} \right) }{Z \times (m + n)}
\end{eqnarray}
was calculated. Here, $Z$ is defined as in Table \ref{AuN's allstructures}, $E_{\text{atom}}^{\text{Au}}$ and $E_{\text{atom}}^{\text{N}}$ are the energies of the spin-polarized non-spherical isolated Au and N atoms, respectively, $E_{\text{solid}}^{\text{Au}_{m}\text{N}_{n}}$ are the bulk cohesive energies calculated by VASP with respect to spherical non spin-polarized isolated atoms, and $m,n = 1,2 \text{ or } 3$ are the stoichiometric ratios.

The calculated $E_{\text{coh}}^{\text{Au}_{m}\text{N}_{n}}$ versus volume per atom $V$ were least-squares-fitted \cite{eos_f90_code} with a 3rd-order Birch-Murnaghan equation of state (EOS)\cite{BM_3rd_eos}. From the fit we obtain the equilibrium cohesive energy $E_{0}$, the equilibrium volume $V_{0}$, the equilibrium bulk modulus
\begin{equation}	\label{B_0 eq}
B_{0} = -V \frac{\partial P}{\partial V}\Big|_{V=V_{0}} = -V \frac{\partial^{2} E}{\partial V^{2}}\Big|_{V=V_{0}}	\; ,
\end{equation}
and the pressure derivative of the bulk modulus
\begin{equation}	\label{B^prime eq}
 B^{\prime}_{0} = \frac{\partial B}{\partial P} \Big|_{P=0} = \frac{1}{B_{0}} \left(  V \frac{\partial}{\partial V} (V \frac{\partial^{2} E}{\partial V^{2}}) \right)  \Big|_{V=V_{0}} 	\; .
\end{equation}
%
\subsection{Formation Energy Calculations}		\label{Formation Energy Calculations}
An important measure of relative stability, beside cohesive energy, is the formation energy $E_{\text{f}}$. Assuming the following chemical reaction between the solid Au(\textit{fcc}) metal and the gaseous N$_{2}$
\begin{eqnarray} \label{E_f reaction}
m \text{Au}^{\text{solid}} + \frac{n}{2} \text{N}_2^{\text{gas}} \rightleftharpoons \text{Au}_m\text{N}_n^{\text{solid}}	\; ,
\end{eqnarray}
$E_{\text{f}}$ of the solid Au$_m$N$_n$ can be obtained from (see Eq. \ref{E_coh equation} for definitions of the quantities):
\begin{align} \label{formation energy equation}
E_{\text{f}}(\text{Au}_m\text{N}_n^{\text{solid}}) =   E_\text{coh}(\text{Au}_m\text{N}_n^{\text{solid}}) \quad \quad \quad \quad \quad \quad &
\nonumber \\
- \frac{  m E_\text{coh}(\text{Au}^{\text{solid}}) + \frac{n}{2} E_\text{coh}(\text{N}_2^{\text{gas}})}{m + n} &		\; .
\end{align}
 We found \footnote{For details on how these properties were calculated, readers are referred to Ref. \onlinecite{Suleiman_PhD_arxiv_copper_nitrides_article}.} the equilibrium cohesive energy of the molecular nitrogen $E_\text{coh}(\text{N}_2^{\text{gas}})$ and its N--N bond length to be $-5.196 \; eV/\text{atom}$ and $1.113 \; \text{\AA}$, respectively. The ground-state cohesive energy $E_\text{coh}(\text{Au}^{\text{solid}})$ and other equilibrium properties of the elemental gold in its standard A1 structure \cite{Wyckoff,Structure_of_Materials,Handbook_of_Mineralogy} are given in Table \ref{table: gold_nitrides_equilibrium_structural_properties}.
%
\subsection{GWA Calculations}	\label{GWA Calculations}
In order to obtain quantitatively accurate optical spectra of Au$_3$N(D0$_9$), it is required to go beyond the realm of traditional DFT \cite{PAW_optics}. A practical method is the so-called GW approach. In this technique, which is provided by many-body perturbation theory (MBPT), a system of quasi-particle (QP) equations \cite{GWA_and_QP_review_1999,Kohanoff,JudithThesis2008}
\begin{eqnarray}	\label{QP equations}
\begin{split}
  \Bigg \{ - \frac{\hbar^{2}} {2m}  \nabla^{2} + & \int d\mathbf{r}^{\prime} \frac{n(\mathbf{r}^{\prime})}{|\mathbf{r}-\mathbf{r}^{\prime}|} + V_{ext}(\mathbf{r}) \Bigg \} \psi_{i,\mathbf{k}}^{QP}(\mathbf{r})  \\  +& \int d\mathbf{r}^{\prime} \Sigma(\mathbf{r},\mathbf{r}^{\prime};\epsilon_{i,\mathbf{k}}^{QP})  \psi_{i,\mathbf{k}}^{QP}(\mathbf{r}^{\prime}) = \epsilon_{i,\mathbf{k}}^{QP} \psi_{i,\mathbf{k}}^{QP}(\mathbf{r})
\end{split}
\end{eqnarray}
is to be solved; where wave functions $\psi_{i,\mathbf{k}}^{QP}(\mathbf{r})$ are taken from the DFT calculations. However, this approach is computationally demanding, and one had to use less dense meshes of $\mathbf{k}$-points, $10 \times 10 \times 10$, while keeping $E_{\text{cut}}$ at $600 \; eV$.

All static and dynamic exchange and correlation effects, including those neglected in the DFT-GGA reference system, are contained in the so-called self-energy $\Sigma(\mathbf{r},\mathbf{r}^{\prime};\epsilon_{i,\mathbf{k}}^{QP})$. Writing $\Sigma$ in terms of the Green's function $G$ and the frequency-dependent screened Coulomb interaction $W$ as
\begin{eqnarray}	\label{GW self-energy}
\begin{split}
\Sigma_{GW} = j \int d\epsilon^{\prime} G(\mathbf{r},\mathbf{r}^{\prime};\epsilon,\epsilon^{\prime}) W(\mathbf{r},\mathbf{r}^{\prime};\epsilon),
\end{split}
\end{eqnarray}
gives rise to the term GW approximation (GWA). $W$ and the bare Coulomb interaction $v$ are related via
\begin{eqnarray}
\begin{split}
W(\mathbf{r},\mathbf{r}^{\prime};\epsilon) = j \int d\mathbf{r}_{1} \varepsilon^{-1}(\mathbf{r},\mathbf{r}_{1};\epsilon)v(\mathbf{r}_{1},\mathbf{r}^{\prime}) \; ,
\end{split}
\end{eqnarray}
where the dielectric Cartesian tensor $\varepsilon$ (in this case is diagonal and isotropic because of the cubic nature of Au$_3$N(D0$_9$)) is calculated within the random phase approximation (RPA).

Following the so-called $GW_{0}$ routine, the QP eigenvalues
\begin{eqnarray} \label{QP eigenvalues}
\begin{split}
\epsilon_{i,\mathbf{k}}^{QP} = \text{Re}  \left( 
\left\langle \psi_{i,\mathbf{k}}^{QP} \middle| 
H_{\text{KS}} - V_{XC} + \Sigma_{GW_{0}}
\middle|  \psi_{i,\mathbf{k}}^{QP} \right\rangle  	\right)
\end{split}
\end{eqnarray}
were updated four times in the calculations of $G$, while $W$ was kept at the DFT-RPA level. Using the updated QP eigenvalues, $\varepsilon$ was recalculated after the execution of the last iteration \cite{Kohanoff,JudithThesis2008,VASPguide}.
%
\subsection{Optical Spectra Calculations}	\label{Optical Spectra Calculations}
Assuming orientation of the Au$_3$N(D0$_9$) crystal surface parallel to the optical axis, it is straightforward then to derive all the desired frequency-dependent optical spectra such as
 absorption coefficient $\alpha\left(\omega\right)$,
 refractive index $n\left(\omega\right)$,
 energy-loss $L\left(\omega\right)$ and
 reflectivity $R\left(\omega\right)$:
\begin{align}
\alpha\left(\omega\right) &= \sqrt{2} \omega  \left(  \left[  \varepsilon_{\text{re}}^{2}\left(\omega\right) + \varepsilon_{\text{im}}^{2}\left(\omega\right) \right]^{\frac{1}{2}}  - \varepsilon_{\text{re}}\left(\omega\right) \right)^{\frac{1}{2}} 		\label{alpha(omega)}\\
n\left(\omega\right) &= \frac{1}{\sqrt{2}} \left(  \left[  \varepsilon_{\text{re}}^{2}\left(\omega\right) + \varepsilon_{\text{im}}^{2}\left(\omega\right) \right]^{\frac{1}{2}} + \varepsilon_{\text{re}}\left(\omega\right) \right)^{\frac{1}{2}} 			\label{n(omega)}\\
L\left(\omega\right) &= \frac{\varepsilon_{\text{im}}\left(\omega\right)}{\varepsilon_{\text{re}}^{2}\left(\omega\right) + \varepsilon_{\text{im}}^{2}\left(\omega\right)} 			\label{L(omega)}\\
R\left(\omega\right) &= \left| \frac{\left[  \varepsilon_{\text{re}}\left(\omega\right) + j \varepsilon_{\text{im}}\left(\omega\right)  \right]^{\frac{1}{2}} - 1}{\left[  \varepsilon_{\text{re}}\left(\omega\right) + j \varepsilon_{\text{im}}\left(\omega\right)  \right]^{\frac{1}{2}} + 1} \right| ^{2}		\label{R(omega)}
\end{align}
 from the real $\varepsilon_{\text{re}}(\omega)$ and the imaginary $\varepsilon_{\text{im}}(\omega)$ parts of the macroscopic $\varepsilon_{\text{RPA}}(\omega)$ \cite{Fox,dressel2002electrodynamics,Ch9_in_Handbook_of_Optics_2010}.

It should be emphasized here that to obtain more accurate optical spectra (that is, more accurate positions and amplitudes of the characteristic peaks), one should solve the so-called Bethe-Salpeter equation, the equation of motion of the two-body Green function $G_2$, in order to include the electron-hole excitations \cite{DFT_GW_BSE_Electron-hole_excitations_and_optical_spectra_from_first_principles_2000}.
%
\section{Results}	\label{Results}
\begin{figure}[!]
\includegraphics[width=0.45\textwidth]{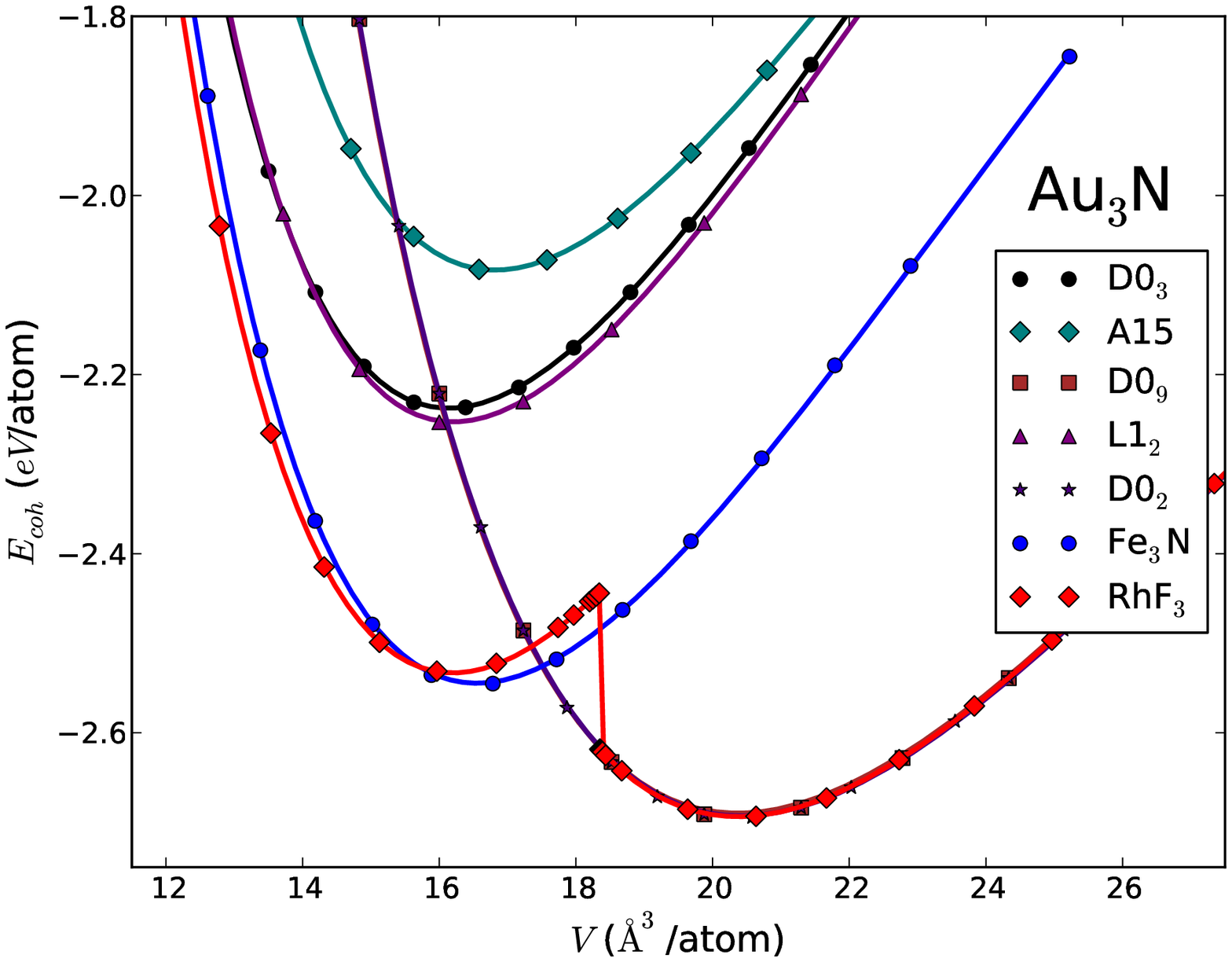}
\caption{\label{Au3N1_ev_EOS}(Color online.) Cohesive energy $E_{\text{coh}} (eV/\text{atom})$ versus volume $V$ (\AA$^{3}$/\text{atom}) for Au$_3$N in seven different structural phases.}
\end{figure}

\begin{figure}[!]
\includegraphics[width=0.45\textwidth]{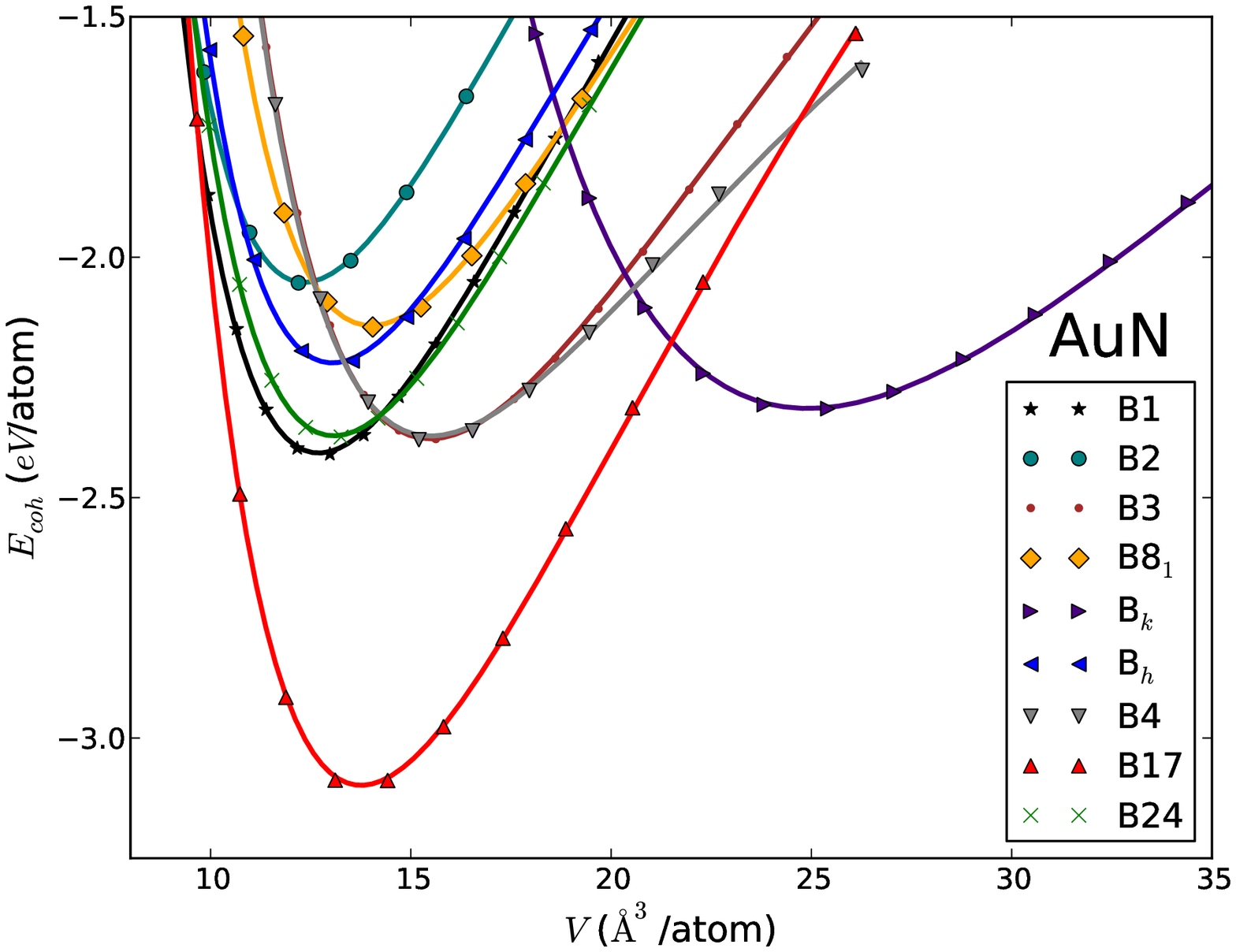}
\caption{\label{Au1N1_ev_EOS}(Color online.) Cohesive energy $E_{\text{coh}} (eV/\text{atom})$ versus volume $V$ (\AA$^{3}$/\text{atom}) for AuN in nine different structural phases.}
\end{figure}

\begin{figure}[!]
\includegraphics[width=0.45\textwidth]{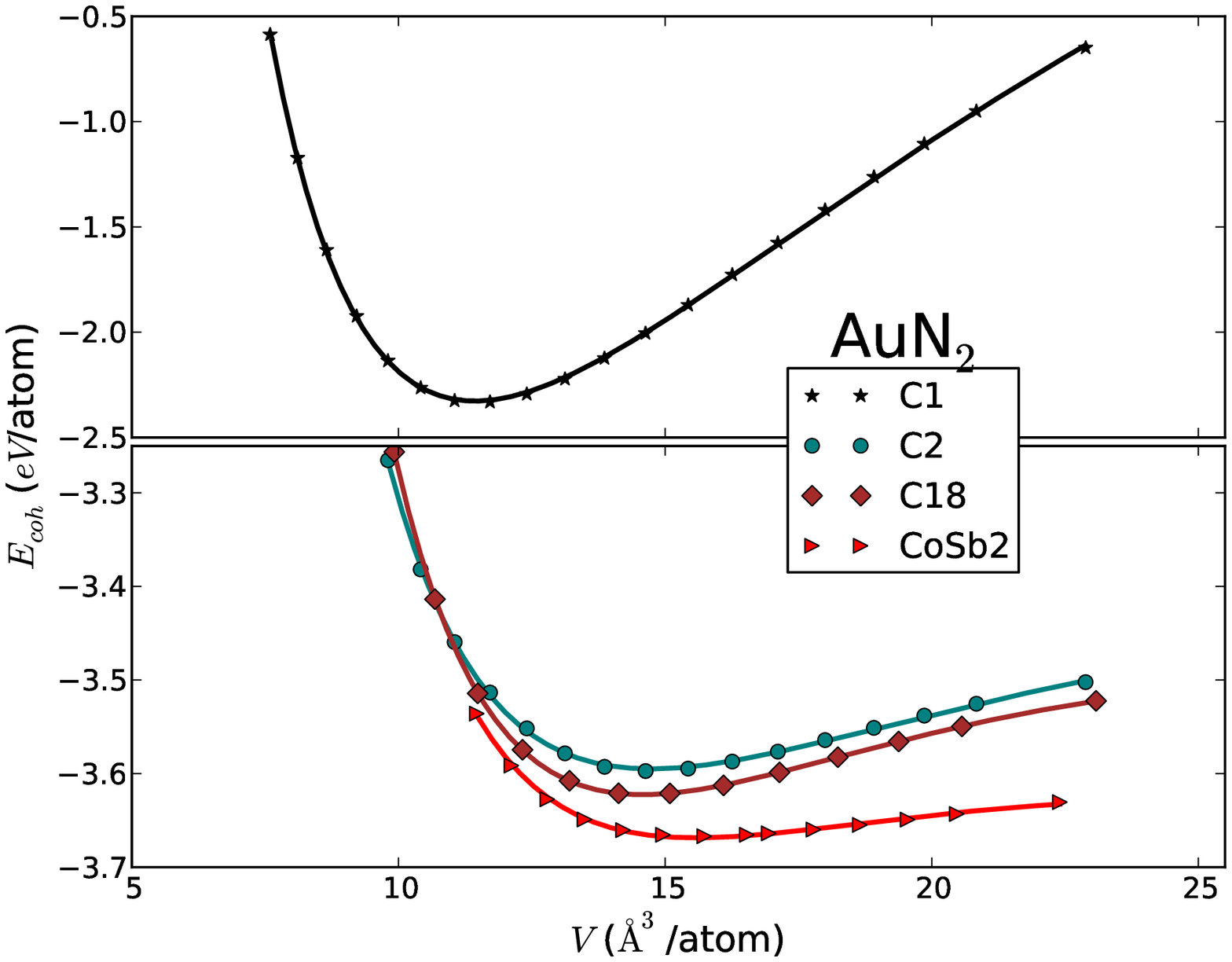}
\caption{\label{Au1N2_ev_EOS}(Color online.) Cohesive energy $E_{coh} (eV/\text{atom})$ versus volume $V (\AA^{3}/\text{atom})$ for AuN$_2$ in four different structural phases.}
\end{figure}

Cohesive energy $E_{\text{coh}}$ versus volume $V_{0}$ equation of state (EOS) for the considered modifications of Au$_3$N, AuN and AuN$_2$ are displayed graphically in Figs. \ref{Au3N1_ev_EOS}, \ref{Au1N1_ev_EOS} and \ref{Au1N2_ev_EOS}, respectively. The corresponding calculated  equilibrium structural, energetic and mechanical properties of these twenty phases and of Au(A1) are presented in Table \ref{table: gold_nitrides_equilibrium_structural_properties}. Modifications in this table are ordered in the same way as in Table \ref{AuN's allstructures}. Whenever possible, our results are compared with experiment and with previous calculations. In the latter case, the calculations methods and the $XC$ functionals are indicated in the Table footnotes.

%
\begin{figure*}
\includegraphics[width=0.95\textwidth]{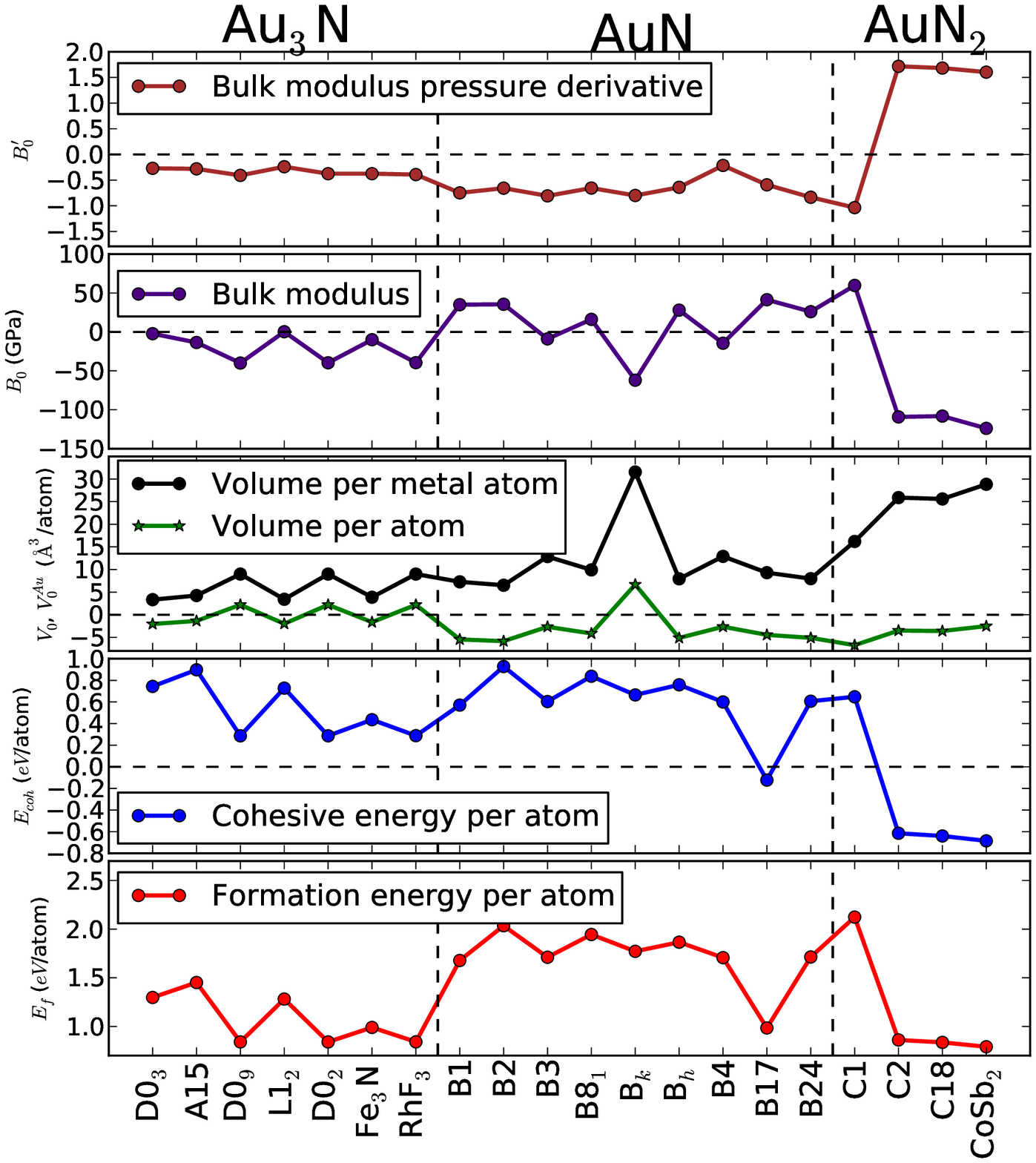}
\caption{\label{figure: gold_nitrides_equilibrium_properties}(Color online.) Calculated equilibrium properties of the twenty studied phases of gold nitrides. All quantities are given relative to the corresponding ones of the \textit{fcc} crystalline elemental gold given in the first row of Table \ref{table: gold_nitrides_equilibrium_structural_properties}.}
\end{figure*}

%
\begin{table*}
\caption{\label{table: gold_nitrides_equilibrium_structural_properties}Equilibrium properties of Au(A1) and of the twenty studied phases of Au$_{3}$N, AuN and AuN$_2$\textbf{:} Lattice constants [$a(\text{\AA})$, $b(\text{\AA})$, $c(\text{\AA})$, $\alpha(^{\circ})$ and $\beta(^{\circ})$], volume $V_{0}(\text{\AA}^{3}/$atom$)$, cohesive energy $E_{\text{coh}} (eV/$atom$)$, bulk modulus $B_{0} (GPa)$, bulk modulus pressure derivative $B_{0}^{\prime}$ and energy of formation $E_{\text{f}}(eV/\text{atom})$. Our obtained values (\textit{Pres.}) are compared to experimentally reported ones (\textit{Expt.}) and to previous calculations (\textit{Comp.}).}
\resizebox{1.0\textwidth}{!}{
\begin{tabular}{lllllllllll}
\hline		
\textbf{Structure}	&		& $a(\AA)$		   & $b(\AA)$			& $c(\AA)$		  & $\alpha(^{\circ})$ or $\beta(^{\circ})$		& $V_{0} (\AA^{3}/$atom$)$   & $E_{\text{coh}}(eV/\text{atom})$	& $B_{0}(\text{GPa})$		& $B_{0}^{\prime}$ & $E_{\text{f}}(eV/\text{atom})$\\
\hline \hline 
					\multicolumn{11}{c}{\textbf{Au}} \\
\hline 
\multirow{4}{*}{\textbf{A1}}    & \textit{Pres.} & $4.174$     &--         & --   & --    & $18.18$  &  $-2.982$   & $135.363$  & $5.926$ 	&  \\
        & \textit{Exp.}& $4.0782$\footnotemark[1] & --    & --       & --            &                           &  $-3.81$\footnotemark[2]    &  $173.2$\footnotemark[2], $173$\footnotemark[3]   & $6.29$\footnotemark[4]	  & \\
        & \textit{Comp.} &  $4.06$\footnotemark[5]\textsuperscript{,}\footnotemark[6]    & --    & --                       & --                        &     &  $-4.38$\footnotemark[7], $-3.17$\footnotemark[8],   &    $187$\footnotemark[5], $205$\footnotemark[6]     &  $4.68$\footnotemark[9], $6.00$\footnotemark[10], & \\

        &  &      & --    & --             & --   &         &   $-3.19$\footnotemark[12]       &        &    $5.23$\footnotemark[11]     &    \\

\hline 
					\multicolumn{11}{c}{\textbf{Au$_3$N}} \\
\hline 
\multirow{1}{*}{\textbf{D0$_3$}}        & \textit{Pres.}&  $6.368$  & --    & --                       & --                        &  $16.14$  & $-2.238$     &  $133.110$  & $5.656$  &  $1.297$ \\
	

\hline 
\multirow{1}{*}{\textbf{A15}}           & \textit{Pres.}& $5.124$ & --    & --                       & --                        &  $16.82$    &  $-2.084$ & $121.792$   &  $5.645$  & $1.451$	\\
	

\hline 
\multirow{2}{*}{\textbf{D0$_9$}}        & \textit{Pres.}&  $4.336$ & --    & --                       & --                        &   $20.38$  &   $-2.695$    &  $95.370$  & $5.518$  & $0.840$ \\
	
         								& \textit{Comp.}&  $4.239$\footnotemark[13] & --    & --                       & --                        &                           &                   &        &   & \\

\hline 
\multirow{1}{*}{\textbf{L1$_2$}}        & \textit{Pres.}& $ 4.017$ & --    & --                       & --                        & $16.20$  &  $-2.254$  &  $135.621$  & $5.686$  & $1.281$	\\
	

\hline 
\multirow{1}{*}{\textbf{D0$_2$}}        & \textit{Pres.}& $8.672$  & --    & --                       & --                        & $20.38$  & $-2.695$   & $95.692$  & $5.551$  & $0.840$		\\
	

\hline 
\multirow{1}{*}{\textbf{$\epsilon$-Fe$_3$N}}& \textit{Pres.}& $5.473$ & --    & $5.100$       & --                        &  $16.54$   &   $-2.546$ &  $125.363$ &  $5.551$  & $0.989$ \\
	

\hline 
\multirow{1}{*}{\textbf{RhF$_3$}}       & \textit{Pres.}&  $6.075$  & --    & --    & $\alpha = 61.269$                        &  $20.38$   &   $-2.694$   &   $95.859$ & $5.534$  & $0.841$ \\
	

\hline 
					\multicolumn{11}{c}{\textbf{AuN}} \\
\hline 
\multirow{1}{*}{\textbf{B1}}   		& \textit{Pres.}& $4.670$   & --    & --                       & --                        & $12.73$ &  $-2.411$  &  $170.385$  & $5.178$   & $1.678$	\\
	

\hline 
\multirow{1}{*}{\textbf{B2}}   		& \textit{Pres.}& $2.912$   & --    & --                       & --                        & $12.35$  & $-2.054$  &  $170.874$ & $5.269$  &  $2.035$	\\
	

\hline 
\multirow{1}{*}{\textbf{B3}}   		& \textit{Pres.}&  $4.989$  & --    & --                       & --                        & $15.52$  &  $-2.378$  &  $126.414$ & $5.119$  &  $1.711$	\\
	

\hline 
\multirow{1}{*}{\textbf{B8$_{1}$}}   	& \textit{Pres.}	&  $3.600$    & --    & $5.007$                       & --      & $14.05$  &  $-2.144$ &  $ 151.504$  & $5.271$  & $1.945$	\\
	

\hline 
\multirow{1}{*}{\textbf{B$_{\text{k}}$}}& \textit{Pres.}	& $3.508$ & --    & $9.332$                       & --                        &  $24.86$   &  $-2.317$  &  $73.343$  &  $5.126$  & $1.772$	\\
	

\hline 
\multirow{1}{*}{\textbf{B$_{\text{h}}$}}& \textit{Pres.}	&  $3.138$  & --    & $3.063$                       & --     &  $13.06$   & $-2.223$   & $163.369$  & $5.285$ & 	$1.866$	\\
	

\hline 
\multirow{1}{*}{\textbf{B4}}   		& \textit{Pres.}	& $3.526$  & --    & $5.774$        & --                        &   $15.54$  &  $-2.382$  & $120.842$   & $5.711$	& 	$1.707$	\\


\hline 
\multirow{1}{*}{\textbf{B17}}   	& \textit{Pres.}	&  $3.149$   & --    & $5.543$                       & --                        &   $13.74$  &    $-3.105$    & $176.760$  & $5.334$   & $	0.984$		\\


\hline 
\multirow{1}{*}{\textbf{B24}}   	& \textit{Pres.}	& $4.380$ & $4.647$    & $5.141$        & --                        &  $13.08$ & $-2.375$  & $161.383$  & $5.092$  & $1.714$	\\

\hline 
					\multicolumn{11}{c}{\textbf{AuN$_2$}} \\
\hline 
\multirow{3}{*}{\textbf{C1}}            & \textit{Pres.}	& $5.162$  & --    & --                       & --                        &  $11.46$   &   $-2.334$  & $195.138$  & $4.890$   & $2.124$	\\

        & \multirow{2}{*}{\textit{Comp.}} & $5.035$\footnotemark[14]     & --    & --                       & --                        & 	 &          & $246$\footnotemark[14] &   & \\
        &  & $5.144$\footnotemark[15]     & --    & --                       & --   &  &       &  $198$\footnotemark[15]  &   & \\

\hline 
\multirow{3}{*}{\textbf{C2}}            & \textit{Pres.}	& $5.607$ & --    & --                       & --                        &  $14.69$  & $-3.597$  & $26.129$  & $7.643$  & $0.861$	\\

        & \multirow{2}{*}{\textit{Comp.}} & $5.471$\footnotemark[16] & --    & --                       & --                        &                           &                   & $41$\footnotemark[16] &   & $0.727$\footnotemark[16] \\

        &                                 & $5.157$\footnotemark[17] & --    & --                       & --                        &                           &                   & $126$\footnotemark[17] &   & \\

\hline 
\multirow{3}{*}{\textbf{C18}}           & \textit{Pres.}	&  $3.467$  & $4.549$    & $5.551$      & --                        &  $14.59$ &  $-3.622$  & $27.178$  & $7.609$  & $0.836$ \\

        & \multirow{2}{*}{\textit{Comp.}} & $6.160$\footnotemark[16] & $5.013$\footnotemark[16]    & $2.936$\footnotemark[16]                       & --                        &                           &                   & $27$\footnotemark[16] &   & $0.554$\footnotemark[16] \\

        &                                 & $5.410$\footnotemark[17] & $4.938$\footnotemark[17] & $2.874$\footnotemark[17]    & --                        &                           &                   & $57$\footnotemark[17]     &   & \\

\hline 
\multirow{3}{*}{\textbf{CoSb$_2$}}      & \textit{Pres.}	&  $6.219$ & $5.882$    & $10.679$  & $\beta = 151.225$                        &  $15.67$  & $-3.667$   & $11.430$  & $7.529$  & 	$0.791$	\\

        & \multirow{2}{*}{\textit{Comp.}} & $8.149$\footnotemark[16] & $5.350$\footnotemark[16]    & $5.361$\footnotemark[16]        & $131.09$\footnotemark[16]  &                           &                   & $36$\footnotemark[16]        &   & $0.529$\footnotemark[16] \\

        &                                 & $7.715$\footnotemark[17] & $5.215$\footnotemark[17]    & $5.172$\footnotemark[17]     & $132.11$\footnotemark[17]  &                           &                   & $81$\footnotemark[17]        &   & \\

\hline \hline 
\end{tabular}
}       

\footnotetext[1]{Ref. \cite{Jerry_1974}. This is an average of 40 experimental values, at $20 \; \celsius$, with a deviation: $\pm 0.0002$ \AA.}
\footnotetext[2]{Ref. \onlinecite{Kittel}. Cohesive energies are given at $0 \; K$ and $1 \text{ atm} = 0.00010 \text{GPa}$; while bulk moduli are given at room temperature.}
\footnotetext[3]{Ref. (25) in \onlinecite{B_prime_1997_theory_comp_n_exp}: at room temperature.}
\footnotetext[4]{See Refs. (8)--(11) in \onlinecite{B_prime_1997_theory_comp_n_exp}.}
\footnotetext[5]{Ref. \onlinecite{elemental_metals_1996_comp}. LAPW-TB.}
\footnotetext[6]{Ref. \onlinecite{elemental_metals_1996_comp}. LAPW-LDA.}
\footnotetext[7]{Ref. \onlinecite{elemental_metals_2008_comp}: PAW-LDA.}
\footnotetext[8]{Ref. \onlinecite{elemental_metals_2008_comp}: PAW-PW91.}
\footnotetext[9]{Ref. \onlinecite{B_prime_1997_theory_comp_n_exp}: using the so-called method of transition metal pseudopotential theory; a modified form of a method proposed by Wills and Harrison to represent the effective interatomic interaction.}
\footnotetext[10]{Ref. \onlinecite{B_prime_1997_theory_comp_n_exp}: using a semi-empirical estimate based on the calculation of the slope of the shock velocity \textit{vs.} particle velocity curves obtained from the dynamic high-pressure experiments. The given values are estimated at $\sim 298 \; K$.}
\footnotetext[11]{Ref. \onlinecite{B_prime_1997_theory_comp_n_exp}: using a semi-empirical method in which the experimental static $P-V$ data are fitted to an EOS form where $B_{0}$ and $B_{0}^{\prime}$ are adjustable parameters. The given values are estimated at $\sim 298 \; K$.}
\footnotetext[12]{Ref. \onlinecite{elemental_metals_2008_comp}: PAW-PBE.}

\footnotetext[13]{Ref. \onlinecite{PhysRevB.70.045414}: Using the AIMPRO code, in which a Gaussian orbital basis set is used with the separable dual-space pseudopotentials of Hutter \textit{et al}.}
\footnotetext[14]{Ref. \onlinecite{AgN2_AuN2_PtN2_2005_comp}: Using the full-potential linearized augmented plane waves (LAPW) method within LDA.}
\footnotetext[15]{Ref. \onlinecite{AgN2_AuN2_PtN2_2005_comp}: Using the full-potential linearized augmented plane waves (LAPW) method within GGA(PBE).}

\footnotetext[16]{Ref. \onlinecite{AuN2_and_PtN2_2010_comp}: using Vanderbilt USPPs within GGA(PBE). $B_0$'s were calculated from the elastic constants. $E_\text{coh}(\text{N}_2^{\text{solid}})$ was used in Eq. \ref{formation energy equation} instead of $E_\text{coh}(\text{N}_2^{\text{gas}})$.}
\footnotetext[17]{Ref. \onlinecite{AuN2_and_PtN2_2010_comp}: using Vanderbilt USPPs within LDA. $B_0$'s were calculated from the elastic constants.}

\end{table*}
To compare and to deeper analyze the obtained equilibrium properties of the three stoichiometries series with respect to one another, the calculated equilibrium properties are depicted graphically in Fig. \ref{figure: gold_nitrides_equilibrium_properties}. All quantities in this figure are given relative to the corresponding ones of Au(A1) given in Table \ref{table: gold_nitrides_equilibrium_structural_properties}. In this way, one will be able to investigate the effect of nitridation on the parent crystalline Au as well \footnote{In Table \ref{table: gold_nitrides_equilibrium_structural_properties}, our computed properties of the elemental Au are compared with experiment and with previous calculations as well. This may benchmark the accuracy of the rest of our calculations.}.

%
\subsection{EOS and Relative Stabilities}	\label{subsection: AuN's: EOS and Relative Stabilities}
Fig. \ref{Au3N1_ev_EOS} shows that the most stable phases in the studied Au$_{3}$N series are D0$_9$, D0$_2$ and RhF$_3$. From the figure, it is also clear that the $E_{\text{coh}}(V)$ curves of these three phases are almost identical around their equilibria. In fact the D0$_9$ curve can hardly be seen in the whole range (compare with Fig. 1(c) in Ref. \onlinecite{Suleiman_PhD_SAIP2012_gold_nitrides_article}). We found the same behavior in the EOS to be true for Ag$_3$N in the same phases (see Ref. \onlinecite{Suleiman_PhD_arxiv_silver_nitrides_article}), and we traced back this behavior to the structural relationships between these three phases (For more details, see Ref. \onlinecite{Suleiman_PhD_arxiv_copper_nitrides_article} and references therein). As can readily be seen from Fig. \ref{figure: gold_nitrides_equilibrium_properties} and Table \ref{table: gold_nitrides_equilibrium_structural_properties}, these structural relations have manifested themselves in  all the presented structural, energetic and mechanical properties, giving rise to almost identical values. Therefore, one may conclude that, if one phase is synthesizable, the three phases may co-exist during the Au$_3$N synthesis process.

Noting that Cu, Ag and Au share the same group in the periodic table of elements, it may be worth to mention here that D0$_{9}$ structure is known to be the structure of the synthesized Cu$_{3}$N \cite{Cu3N_1989_exp,Cu3N_2006_exp,Cu3N_Ni3N_1993_exp} and, as mentioned above, we found it to be the most stable structure of Ag$_{3}$N \cite{Suleiman_PhD_arxiv_silver_nitrides_article}.

Assuming that it is the most likely stoichiometry, Krishnamurthy \textit{et al.} \cite{PhysRevB.70.045414} undertook \textit{ab initio} pseudopotential calculations on Au$_3$N and studied all the Au$_{3}$N structures in Table \ref{AuN's allstructures}. Although they found D0$_{9}$ to be the most stable modification in this sub-parameter space, yet, they identified a triclinic crystal structure with $0.25 \; eV/\text{atom}$ lower energy than the D0$_9$. Krishnamurthy and co-workers determined the triclinic phase to be metallic. It must be mentioned here that all the 3:1 structures we have investigated in the present study were taken mainly from the work of Krishnamurthy \textit{et al.} \cite{PhysRevB.70.045414}. However, Krishnamurthy et. al. gave only the \textit{lattice vectors} of their triclinic structure, but no \textit{basis vectors} were given. So, we were not able to properly place the atoms inside the cell they gave. Allowing them to relax, ions keep moving over the potential surface with no sign of a local minimum, and the structure seems to be very \textit{soft}! 

The odd behavior of the EOS curve of Au$_3$N(RhF$_3$) with the existence of two minima (Fig. \ref{Au3N1_ev_EOS}) reveals that the first minimum (the one with higher $E_{\text{coh}}$) is a metastable local minimum on the potential surface that cannot be maintained as the bulk Au$_3$N(RhF$_3$) is decompressed. The potential barrier, represented by the sudden drop of the Au$_3$N(RhF$_3$) curve, at $\sim 18.4 \; eV/\text{atom}$  is due to the change of positions of those metal ions which possess internal degrees of freedom.

Concerning the AuN series, it is evident from Fig. \ref{Au1N1_ev_EOS} that the simple tetragonal structure of cooperite (B17) would be the energetically most stable structure, with $0.694 eV/\text{atom}$ less than B1. This B17 structure was theoretically predicted to be the ground-state structure of CuN \cite{Suleiman_PhD_arxiv_copper_nitrides_article}, AgN \cite{Suleiman_PhD_arxiv_silver_nitrides_article}  and PtN \cite{PtN_2006_comp_B17_structure_important,Suleiman_PhD_arxiv_platinum_nitrides_article}.

Kanoun and Said \cite{CuN_AgN_AuN_2007_comp} studied the $E(V)$ EOS for AuN in the B1, B2, B3 and B4 structures. Within this parameter sub-space, the relative stabilities they arrived at agree in general with ours. However, they predicted that B3 is \textit{always} more stable than B4, while Fig. \ref{Au1N1_ev_EOS} shows that B4 is preferred against B3 only at low pressures.

In the AuN$_{2}$ series, the least symmetric simple CoSb$_{2}$ monoclinc structure is found to be the most stable (Fig. \ref{Au1N2_ev_EOS}). This agrees with the conclusion of Ref. \onlinecite{AuN2_and_PtN2_2010_comp}, where it is suggested that AuN$_{2}$ may be synthesized at extreme conditions (higher pressure and temperature) and/or it may have other Au:N stoichiometric ratios than 1:2.
%
\subsection{Pressure-induced phase transitions}	\label{subsection: AuN's: Pressure-induced phase transitions}
\begin{figure}[!]
\includegraphics[width=0.45\textwidth]{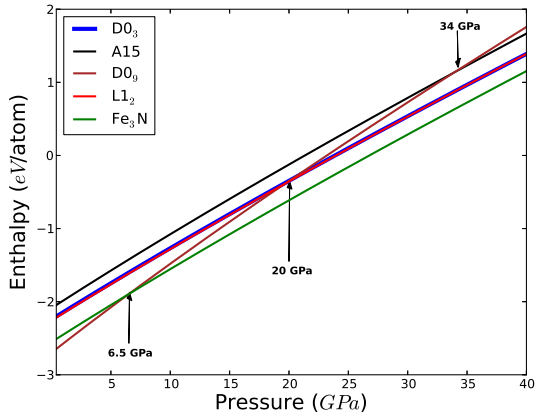}
\caption{\label{Au3N1_hp_EOS}(Color online.) Enthalpy $H$ versus pressure $P$ for Au$_3$N in five structures.}
\end{figure}

Enthalpy-pressure relations of Au$_3$N for five considered structures are displayed in Fig. \ref{Au3N1_hp_EOS}. A point at which enthalpies $H = E_{\text{coh}}(V ) + PV$ of two structures are equal defines the transition pressure $P_{t}$, where transition from the phase with higher enthalpy to the phase with lower enthalpy may occur.

Fig. \ref{Au3N1_hp_EOS} shows that a transition from D0$_{9}$ phase to the Fe$_3$N phase would take place at a very low pressure $\sim 6.5$ GPa; and it is clear that the D0$_{9}$ phase is favourable only at low pressures below $\sim 6.5$ GPa, while the Fe$_3$N hexagonal structure of Ni$_{3}$N is favoured at higher pressures. Fig. \ref{Au3N1_hp_EOS} also reveals that L1$_{2}$ and D0$_3$ phases may co-exist over a wide range of pressure and that they are both favoured over D0$_{9}$ phase at pressures higher than $\sim 20$ GPa, while A15 would be favoured over D0$_{9}$ only at pressures higher than $\sim 34$ GPa.
%
\subsection{Volume per Atom and Lattice Parameters}	\label{subsection: AuN's: Volume per Atom and Lattice Parameters}
The obtained equilibrium lattice parameters and the corresponding volume per atom $V_0$ for the twenty modifications are given in Table \ref{table: gold_nitrides_equilibrium_structural_properties}. The middle sub-window of Fig. \ref{figure: gold_nitrides_equilibrium_properties} shows the $V_0$ values relative to Au(A1). To measure the average distance between two Au ions in the gold nitride crystal, the equilibrium average volume per Au atom ($V_0^{\text{Au}}$), which is simply the ratio of the volume the unit cell to the number of Au atoms in that unit cell, is depicted in the same subwindow.

From the $V_0$ graph in Fig. \ref{figure: gold_nitrides_equilibrium_properties}, it is clear that all AuN and AuN$_2$ modifications, except the open AuN(B$_\text{k}$) phase, have lower $V_0$ than the parent Au(A1); while the Au$_3$N phases tend, on average, not to change the number density of the parent Au(A1). The metal-metal bond length, as represented by the volume per \textit{metal} atom $V_0^{\text{Au}}$, increases (on average) in the direction of increasing nitrogen content and decreasing structural symmetry.

Both trends in $V_0$ and in $V_0^{\text{metal}}$ were found to be true for copper \cite{Suleiman_PhD_arxiv_copper_nitrides_article}, silver \cite{Suleiman_PhD_arxiv_silver_nitrides_article} and platinum \cite{Suleiman_PhD_arxiv_platinum_nitrides_article} nitrides. The trend in $V_0^{\text{metal}}$, however, reveals the fact that in all these nitrides, the introduced N ions displace apart the ions of the host lattice causing longer metal-metal bonds than in the elemental parent metal. This is not readily seen from the $V_0$ curve depicted in the same sub-figure.
%
\subsection{Mechanical Properties}		\label{subsection: AuN's: Bulk Modulus and its Pressure Derivative}
The numerical values of the equilibrium bulk moduli and their pressure derivatives for the twenty modifications are presented in Table \ref{table: gold_nitrides_equilibrium_structural_properties}. The second from top and the top sub-windows of Fig. \ref{figure: gold_nitrides_equilibrium_properties} visualize these values relative to Au(A1).

In the Au$_3$N series, one can see from the second top subfigure of Fig. \ref{figure: gold_nitrides_equilibrium_properties} that less stable phases tend to preserve the $B_0$ value of their parent Au(A1), while the most stable phases (D0$_9$, D0$_2$ and RhF$_3$) posses lower $B_0$ values.

Except B$_{\text{k}}$, AuN modifications and AuN$_2$(C1) tend, on average, to increase the $B_0$ value of their parent Au(A1), with the highest $B_0$ value possessed by the most stable AuN phase: B17.

The last least symmetric structures AuN$_2$(C2, C18 and CoSb$_2$) possess the lowest $B_0$ values among the 20 studied modifications. The $B_0$ values of AuN$_2$'s have the same trend as their corresponding $E_{\text{coh}}$'s and opposite trend as their corresponding $V_0^{\text{Au}}$'s.

From the top subfigure of Fig. \ref{figure: gold_nitrides_equilibrium_properties} one can see that upon application of external pressure, all Au$_3$N and AuN phases and C1 phase tend to lower their $B_0$. The last AuN$_2$ three modifications, however, are more sensitive to external pressure, and their bulk moduli tend to increase under pressure.
%
\subsection{Thermodynamic Stability}		\label{subsection: AuN's: Thermodynamic Stability}
The positive sign of the calculated formation energy $E_{\text{f}}$ (Table \ref{table: gold_nitrides_equilibrium_structural_properties} and their graphical representation in Fig. \ref{figure: gold_nitrides_equilibrium_properties}) means, in principle, that all these modifications are thermodynamically unstable. However, it is common that one obtains positive DFT-calculated $E_{\text{f}}$ for (even the experimentally synthesized) transition-metal nitrides. Moreover, the zero-pressure zero-temperature DFT calculations have to be corrected for the conditions of formation of these nitrides. Another source of this apparent shortcoming stems from the PBE-GGA underestimation of the cohesion in N$_{2}$. We have discussed this point further in Ref. \onlinecite{Suleiman_PhD_arxiv_copper_nitrides_article}. Nevertheless, the presented energies of formation are used as a measure of \textit{relative} thermodynamic stability. That is, the lower the formation energy of any of the phases under consideration, the lower its tendency to dissociate back into its constituent components Au and N$_2$.

Fig. \ref{figure: gold_nitrides_equilibrium_properties} reveals that, within each series and relative to each other, the formation energy of the considered modifications has the same trend as the cohesive energy\footnote{\textit{Surely, this has to be so; since, for a given chemical formula, definitions \ref{E_coh equation} and \ref{formation energy equation} differ only by a constant; that is, the second term in Eq. \ref{formation energy equation}.}}. In other words, all modifications with the same stoichiometric ratios have the same relative stabilities in the formation energy space as in the cohesive energy space. Nonetheless, while Au$_3$N phases tend to have comparable $E_{\text{coh}}$ as the AuN phases, all Au$_3$N modifications have a lower $E_{\text{f}}$ than the AuN ones, except B17. In fact, Fig. \ref{figure: gold_nitrides_equilibrium_properties} indicates that it may be relatively hard to form a 1:1 gold nitride other than B17. Moreover, it is apparent that the tendency of AuN$_2$(C2, C18 and CoSb$_2$) phases to decompose back to Au and N$_2$ is comparable with that of Au$_3$N(D0$_2$, D0$_9$ and RhF$_3$).

Using Vanderbilt USPPs within GGA(PBE), Chen, Tse and Jiang\cite{AuN2_and_PtN2_2010_comp} calculated $E_{\text{f}}$ of C2, C18 and CoSb$_2$ phases. Their results are included in Table \ref{table: gold_nitrides_equilibrium_structural_properties} for comparison. Although they got positive values, as expected, the differences between our obtained values and theirs can be traced back to the fact that they used  $E_{\text{coh}}(\text{N}_2^{\text{solid}})$ in Eq. \ref{formation energy equation} instead of $E_{\text{coh}}(\text{N}_2^{\text{gas}})$. Moreover, the smaller the difference between our obtained values and their obtained equilibrium lattice parameters, the smaller the difference in $E_{\text{f}}$.

Unfortunately, no experimental values of $E_{\text{f}}$ for the synthesized gold nitride phases are available.
%
\subsection{Electronic Properties}		\label{subsection: AuN's: Electronic Properties}
In this subsection, the DFT calculated electronic structure for the most energetically stable phases are shown graphically. In each case, presented information include \textbf{(a)} spin-projected total density of states (TDOS); \textbf{(b)} partial density of states (PDOS) of Au($s, p, d$) orbitals in Au$_3$N; \textbf{(c)} PDOS of N($s, p$) orbitals in Au$_3$N, and \textbf{(d)} band structure along the high-symmetry $\mathbf{k}$-points.

\begin{figure*}
\includegraphics[width=1.0\textwidth]{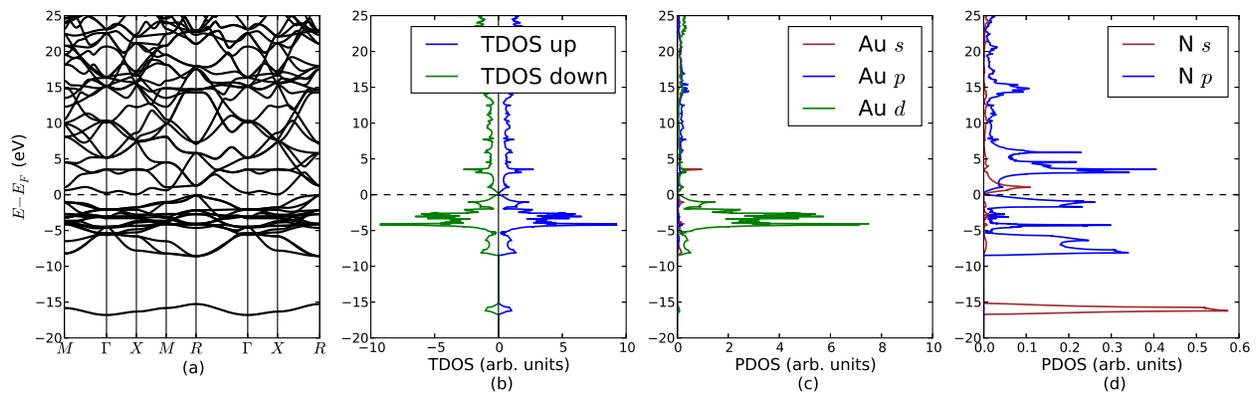}
\caption{\label{Au3N1_D0_9_electronic_structure}(Color online.) DFT calculated electronic structure for Au$_3$N in the D0$_9$ structure:
 \textbf{(a)} band structure along the high-symmetry $\mathbf{k}$-points which are labeled according to Ref. \onlinecite{Bradley}. Their coordinates w.r.t. the reciprocal lattice basis vectors are: $M(0.5, 0.5, 0.0)$, $\Gamma(0.0, 0.0, 0.0 )$, $X(0.0, 0.5, 0.0 )$, $R(0.5, 0.5, 0.5 )$; 
 \textbf{(b)} spin-projected total density of states (TDOS);
 \textbf{(c)} partial density of states (PDOS) of Au($s, p, d$) orbitals in Au$_3$N; and 
 \textbf{(d)} PDOS of N($s, p$) orbitals in Au$_3$N.}
\end{figure*}

\begin{figure*}
\includegraphics[width=1.0\textwidth]{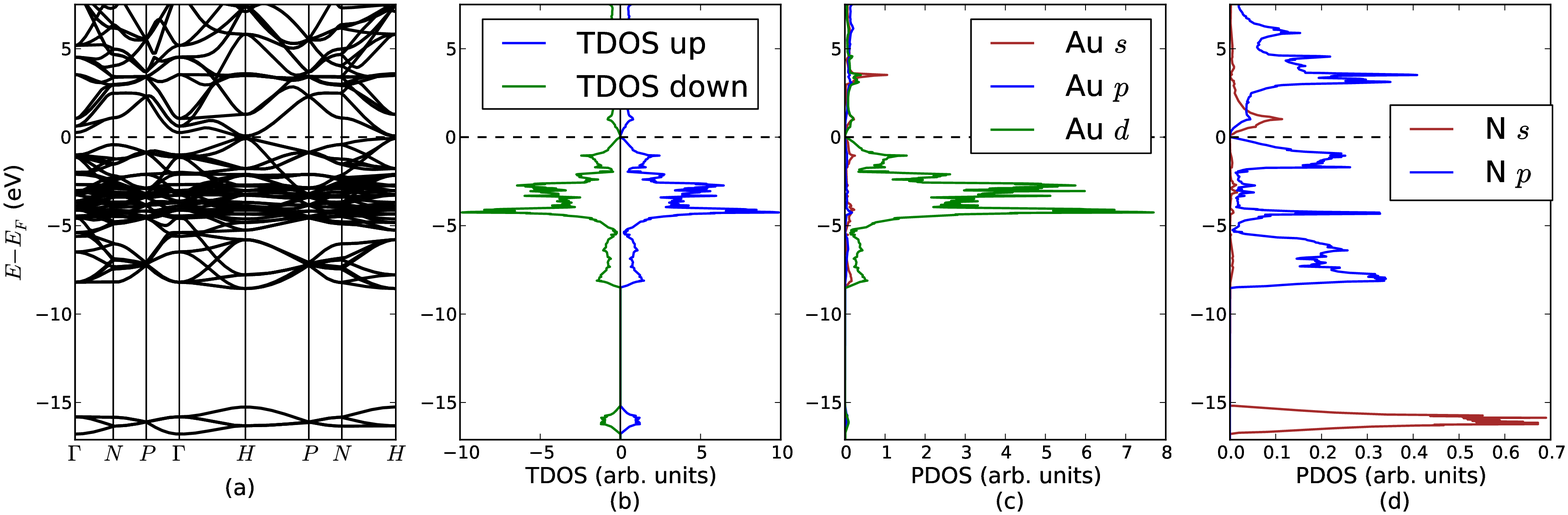}
\caption{\label{Au3N1_D0_2_electronic_structure}(Color online.) DFT calculated electronic structure for Au$_3$N in the D0$_2$ structure:
 \textbf{(a)} band structure along the high-symmetry $\mathbf{k}$-points which are labeled according to Ref. \onlinecite{Bradley}. Their coordinates w.r.t. the reciprocal lattice basis vectors are: $\Gamma(0.0, 0.0, 0.0)$, $N(0.0, 0.0, 0.5)$, $P(0.25, 0.25, 0.25)$, $H(0.5, -.5, 0.5)$;
 \textbf{(b)} spin-projected total density of states (TDOS);
 \textbf{(c)} partial density of states (PDOS) of Au($s, p, d$) orbitals in Au$_3$N; and
 \textbf{(d)} PDOS of N($s, p$) orbitals in Au$_3$N.}
\end{figure*}

\begin{figure*}
\includegraphics[width=1.0\textwidth]{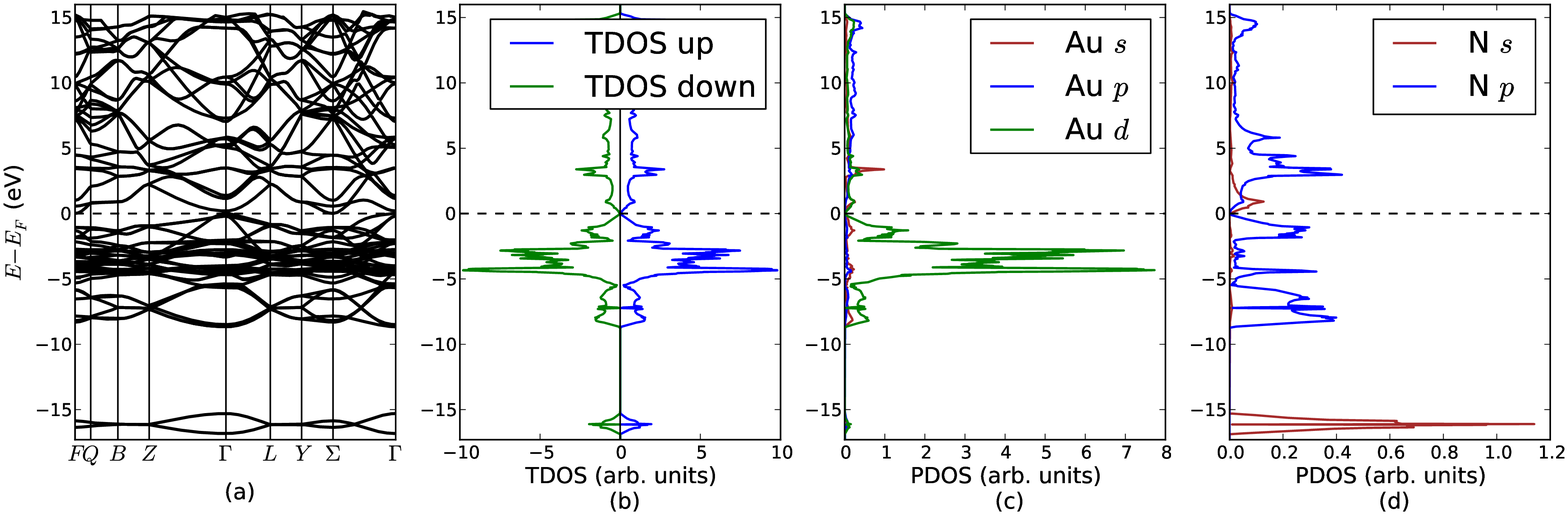}
\caption{\label{Au3N1_RhF_3_electronic_structure}(Color online.) DFT calculated electronic structure for Au$_3$N in the RhF$_3$ structure:
 \textbf{(a)} band structure along the high-symmetry $\mathbf{k}$-points which are labeled according to Ref. \onlinecite{Bradley}. Their coordinates w.r.t. the reciprocal lattice basis vectors are: $F(0.5, 0.5, 0.0)$, $Q(0.375, 0.625, 0.0)$, $B(0.5, 0.75, 0.25)$, $Z(0.5, 0.5, 0.5)$, $\Gamma(0.0,  0.0, 0.0)$, $L(0.0, 0.5, 0.0)$, $Y(0.25, 0.5, -.25)$, $\Sigma(0.0, 0.5, -.5)$;
 \textbf{(b)} spin-projected total density of states (TDOS);
 \textbf{(c)} partial density of states (PDOS) of Au($s, p, d$) orbitals in Au$_3$N; and
 \textbf{(d)} PDOS of N($s, p$) orbitals in Au$_3$N.}
\end{figure*}

\quad \\
Beside D0$_9$ (Fig. \ref{Au3N1_D0_9_electronic_structure}), the equilibrium electronic structure of its two competing phases: D0$_2$ and RhF$_3$, are presented in Figs. \ref{Au3N1_D0_2_electronic_structure} and \ref{Au3N1_RhF_3_electronic_structure}, respectively. This is because the foregoing similarity in their EOS's may reflect in their electronic properties.

%
%

Krishnamurthy \textit{et al.} \cite{PhysRevB.70.045414} predicted Au$_{3}$N(D0$_{9}$) to be an indirect band-gap semiconductor, but they did not give a value. Fig. \ref{Au3N1_D0_9_electronic_structure} shows that it is indeed a semiconductor with an $(R-X)$ indirect DFT band gap of $0.139 \; eV$ GGA value. According to the fact that the produced gold nitrides are metallic, the D0$_{9}$ structure may not be the true candidate for the most likely stoichiometry, Au$_{3}$N.

Fig. \ref{Au3N1_D0_2_electronic_structure} shows that Au$_3$N(D0$_2$) has its CBM at $(H, 0.065\; eV)$, and its VBM is at $(H,-0.073 \; eV)$, resulting in a direct energy band gap at $H$: $E_g = 0.139 \; eV$. This is exactly equal to $E_g$ of Au$_3$N(D0$_9$).

The Fermi surface $E_F$ in Au$_3$N(RhF$_3$) crosses the valence band at $\Gamma$ and the phase seems to have a poor metallic character, since there is only a very narrow width of energy of the unoccupied states above $E_F$ and around $\Gamma$.

A common feature in these three Au$_3$N phases, there is an Au($d$)-N($p$) mixture in the range $\sim \; -8.8 - E_{\mathrm{F}}$ which becomes stronger around $-4.45 \; eV$.

\begin{figure*}
\includegraphics[width=1.0\textwidth]{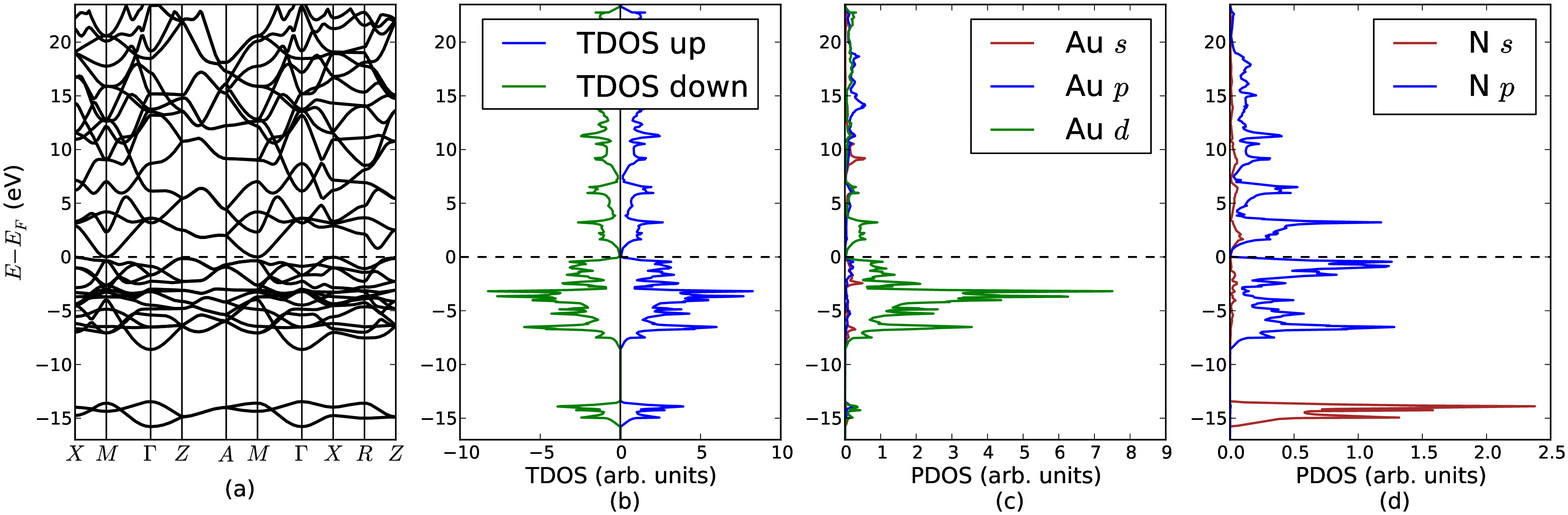}
\caption{\label{Au1N1_B17_electronic_structure}(Color online.) DFT calculated electronic structure for AuN in the B17 structure:
\textbf{(a)} band structure along the high-symmetry $\mathbf{k}$-points which are labeled according to Ref. \onlinecite{Bradley}. Their coordinates w.r.t. the reciprocal lattice basis vectors are: $X (0.0, 0.5, 0.0)$, $M (0.5, 0.5, 0.0)$, $\Gamma (0.0, 0.0, 0.0)$, $Z (0.0, 0.0, 0.5)$, $A (0.5, 0.5, 0.5)$, $R (0.0, 0.5, 0.5)$;
  \textbf{(b)} spin-projected total density of states (TDOS);
  \textbf{(c)} partial density of states (PDOS) of Au($s, p, d$) orbitals in AuN; and
  \textbf{(d)} PDOS of N($s, p$) orbitals in AuN.}
\end{figure*}

Although it might not be clear on the graph, Fig. \ref{Au1N1_B17_electronic_structure}(a) shows that AuN(B17) is a DFT(GGA) indirect band gap semiconductor. With its valence band maximum (VBM) at  $(X,-0.012)$ and its conduction band minimum (CBM) very close to $E_{\text{F}}$ at $(M,0.001 \;  eV)$, AuN(B17) possesses a very narrow band gap of width: $E_g = 0.013 \; eV$.

This insulating feature is in contrast to 
 PdN \cite{Suleiman_PhD_SAIP2012_palladium_nitrides_article},
 PtN \cite{Suleiman_PhD_arxiv_platinum_nitrides_article},
 CuN \cite{Suleiman_PhD_arxiv_copper_nitrides_article}
 and AgN \cite{Suleiman_PhD_arxiv_silver_nitrides_article} which were all found to be metallic in this B17 structure.

\begin{figure*}
\includegraphics[width=1.0\textwidth]{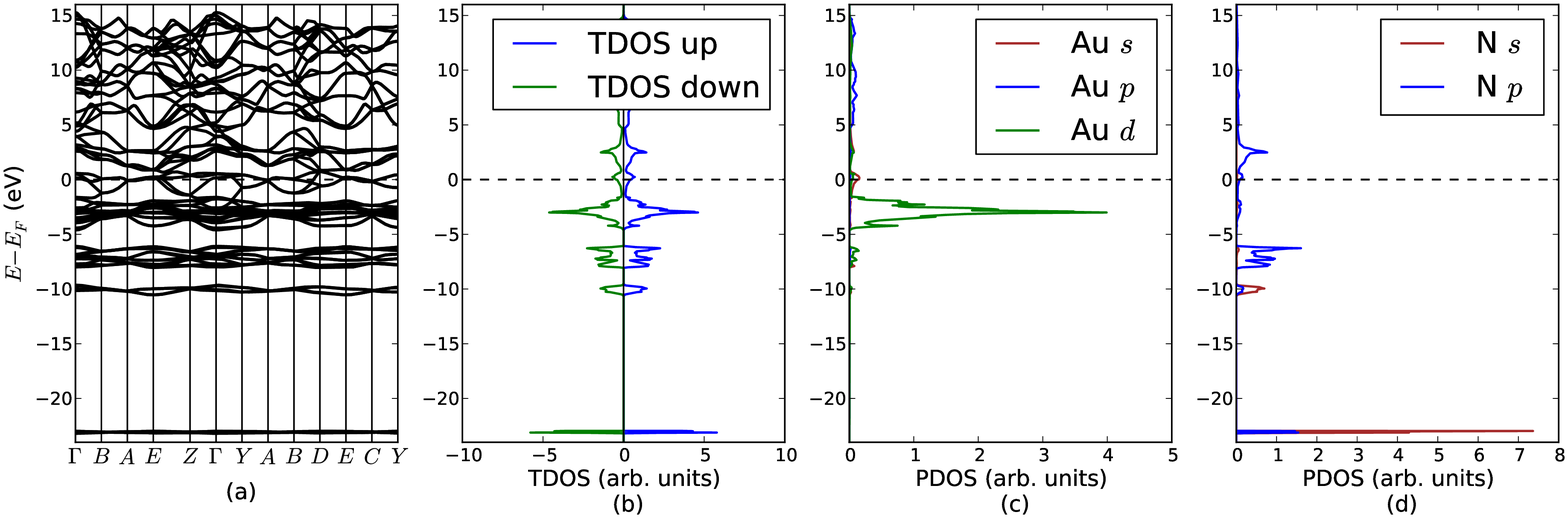}
\caption{\label{Au1N2_CoSb2_electronic_structure}(Color online.) DFT calculated electronic structure for AuN$_2$ in the CoSb$_{2}$ structure:
 \textbf{(a)} band structure along the high-symmetry $\mathbf{k}$-points which are labeled according to Ref. \onlinecite{Bradley}. Their coordinates w.r.t. the reciprocal lattice basis vectors are: $\Gamma( 0.0, 0.0, 0.0)$, $B(-.5   ,  0.0  ,   0.0)$, $A(-.5   ,  0.5  ,   0.0)$, $E(-.5   ,  0.5  ,   0.5)$, $Z(0.0   ,  0.0  ,   0.5)$, $Y(0.0   ,  0.5  ,   0.0)$, $D(-.5   ,  0.0  ,   0.5)$ and $C(0.0   ,  0.5  ,   0.5)$;
 \textbf{(b)} spin-projected total density of states (TDOS);
 \textbf{(c)} partial density of states (PDOS) of Au($s, p, d$) orbitals in AuN$_2$; and
 \textbf{(d)} PDOS of N($s, p$) orbitals in AuN$_2$.}
\end{figure*}

With the Fermi surface crossing many partly occupied bands, it is evident from Fig. \ref{Au1N2_CoSb2_electronic_structure} that AuN$_2$(CoSb$_2$) is a metal.

%
\subsection{Optical Properties} \label{subsection: AuN's: Optical Properties}
\begin{figure*}
\includegraphics[width=1.0\textwidth]{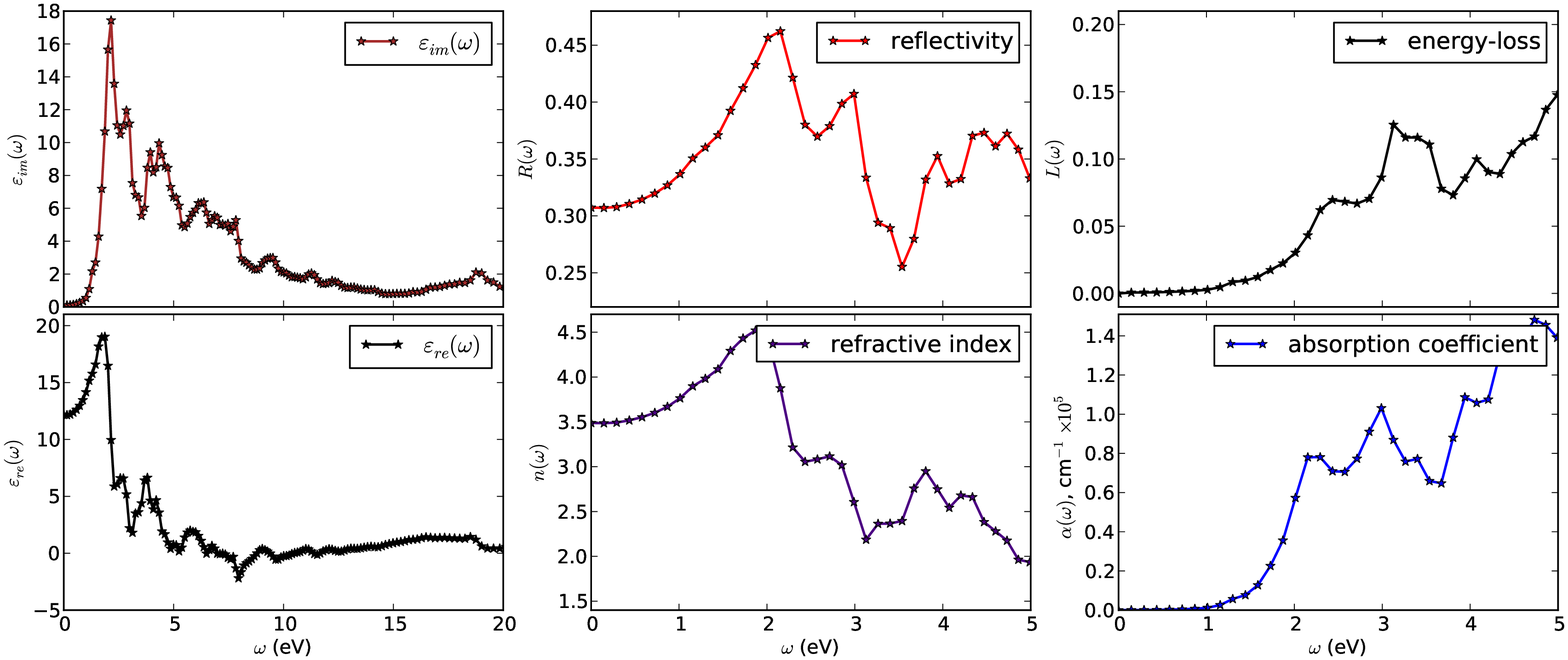}
\caption{\label{Au3N1_D0_9_optical_constants}(Color online.) Normal-incidence frequency-dependent optical spectra of Au$_3$N(D0$_9$) obtained using $GW_{0}$ eigenvalues and Eqs. \ref{R(omega)}-- \ref{alpha(omega)}.} 
\end{figure*}

Within a frequency range that includes the optical region (i.e. the visible spectrum: $[(390 \sim 750 \; \text{nm}) \equiv (3.183 \sim 1.655 \; eV)]$), Fig. \ref{Au3N1_D0_9_optical_constants} displays the real and the imaginary parts of the frequency-dependent dielectric function $\varepsilon_{\text{RPA}}(\omega)$ of Au$_3$N(D0$_9$) and the corresponding derived optical constants. 

It can be seen from the absorption coefficient $\alpha\left(\omega\right)$ spectrum that Au$_{3}$N(D0$_{9}$) posseses a band gap of $\sim 0.9 \; \text{eV}$, since it starts absorbing photons with that energy.  Therefore, it is clear that $GW_{0}$ calculations give a band gap of $\sim 0.9 \; \text{eV}$, which is a significant improvement to the obtained DFT-GGA value. Hence, our presented $\alpha\left(\omega\right)$ spectrum confirms that Au$_{3}$N(D0$_{9}$) would be a semiconductor and that D0$_{9}$ cannot be the true structure of the most likely Au$_{3}$N stoichiometry.
%
\section{Concluding Remarks}	\label{section: AuN's: Concluding Remarks}
We have successfully applied first-principles calculation methods to investigate the structural, stability, electronic and optical properties of Au$_3$N, AuN and AuN$_2$. Within the accuracy of the employed methods, the obtained structural parameters, EOS, $B_{0}$, $B_{0}^{\prime}$ and electronic properties show acceptable agreement with some of the available previous calculations.

Among the studied modifications, we determined metallic (RhF$_3$ and CoSb$_2$) and semiconducting (D0$_9$ and B17) phases.

According to the fact that the produced gold nitride phases are metallic, our DFT-GGA and GW calculations confirmed that D0$_{9}$ structure cannot be the true candidate for the Au$_{3}$N stoichiometry that has been suggested by experimentalists.

From experiment, \textit{ab initio} calculations of Krishnamurthy \textit{et al.} \cite{PhysRevB.70.045414}, and from the present work, one may conclude that if Au$_3$N is the true stoichiometry, it must have a metallic character only at low crystal symmetries: i.e. RhF$_3$ (present work) or a triclinic (Ref. \onlinecite{PhysRevB.70.045414}). However, the better hardness -compared to pure gold- of the synthesized phases\cite{gold_nitride_2005_expt} remains a mystery and may be a property of gold nitride at low dimensions only.

The low symmetry AuN$_2$ phases have far lower cohesive energy than all Au$_3$N, have comparable formation energy with the most favorable Au$_3$N modifications, and their bulk moduli become higher under pressure.
\bigskip
%
\section*{Acknowledgments}
All GW calculations and some DFT calculations were carried out using the infrastructure of the Centre for High Performance Computing (CHPC) in Cape Town. Suleiman would like to thank Dr Mahlaga P. Molepo and Dr Mustafa A. A. Ahmed for correcting misspellings and for their comments. Suleiman would also like to acknowledge the support he received from Wits, DAAD, AIMS and SUST. Many thanks to the ASESMA family, and special thanks to Dr Kris Delaney, Ms Sinead Griffin and Prof Shobhana Narasimhan for their invaluable help and useful discussions. 
%
\bibliography{v4_arXiv_gold_nitrides_article}

\end{document}